# Nonlinear effects in locally-resonant nanostrip phononic metasurface at GHz frequencies

Feng Gao,[1,2, a] Amine Bermak,[1] Sarah Benchabane,[2] Marina Raschetti,[2] and Abdelkrim Khelif[2]

[1] College of Science & Engineering, Hamad Bin Khalifa University, Doha, Qatar
[2] MN2S Department, FEMTO-ST Institute, CNRS, Besancon, France

[a] Author to whom correspondence should be addressed: fgao@hbku.edu.qa

*Abstract*
In this paper, we report on the observation of nonlinear effects in a nanostrip phononic metasurface (NPM) that enable the tuning of resonance frequencies at 1.42 GHz. The NPM resonator made of periodic nanostrip array is fabricated on a lithium niobate substrate. Each of the nanostrip is 250-nm wide and is made of 680-nm-thick $SiO_2$ layer stacking on 50-nm Al metal electrodes. Finite element analysis reveals that the device operates in a vertically polarized (compression) mode with substantial acoustic energy confined in the nanostrips, leading to a local resonance at low acoustic velocity. Due to the nonlinearity, the resonance frequency of the device decreases quadratically with the increase of stimulation power from 10 to 30 dBm. The underlying mechanism of the nonlinearity is found to be the power-dependent coupling of the adjacent nanostrips. This coupling induces softening of the substrate surface region, which reduces the acoustic velocity and hence the bulk radiation. As a result, the quality factor of the NPM resonator is found to improve with the increase of stimulation power. The power-dependent coupling of nanostrips in the NPM resonator demonstrates a reliable method for the realization of nonlinearity in phononic metasurfaces, which would significantly enrich the mechanisms for the manipulation of surface acoustic waves at high frequencies.

Nonlinearity in phononic crystals, acoustic/elastic metamaterials and phononic metasurfaces is a desired effect for the implementation of many useful acoustic phenomena including breathers[1], bifurcation[2], tunability[3], localization[4], and chaos[5]. Phononic nonlinearities were previously exploited in nanomechanical waveguides[3, 6], magnetic lattices[7, 8], surface acoustic wave (SAW) gratings[9] and lamb wave devices[10]. Nevertheless, most of the demonstrations were in the kHz or MHz frequency ranges. Nonlinearities of phononic metasurfaces operating in GHz frequencies are still to be explored.

A good option for achieving nonlinearity in phononic crystal is to build a phononic pillar array whose composing elements are coupled nanomechanical resonators that exhibit nonlinearity. As an example, Midtvedt *et. al* exploited this idea by introducing coupling through an atomically thin graphene membrane transferred to the top of a phononic crystal[11]. But coupling nano or micro-resonators can also be achieved through the substrate. Recently, it was reported that cylindrical pillars very close to each other exhibit substrate-induced coupling that enables phase modulation of the pillars[12]. Utilizing this phenomenon, we built a nanostrip phononic metasurface (NPM) resonator with high-aspect-ratio nanostrips that exhibit substrate coupling induced nonlinearity. The coupling can be controlled by the stimulation power, which results in a power-dependent frequency shift of the NPM resonators.

Fig. 1 shows a diagram of the NPM made of periodic high-aspect-ratio nanostrips on a 128° Y-cut lithium niobate ($LiNbO_3$) substrate. Each of the nanostrip is made of a stack of materials as shown in the insert of Fig. 1. The majority of the stack consists of a 680-nm-thick silicon dioxide layer, serving as the structural material. At the bottom of the stack lies a metal electrode made of 50-nm aluminum (Al) sandwiched by two layers of 5-nm-thick titanium (Ti). Al is the conducting material while the Ti layers are used to promote adhesion. The width of the nanostrip is 250 nm, leading to an aspect ratio of 2.96. Due to the high aspect ratio, a substantial amount of the acoustic energy is stored in the nanostrips, which forms a locally confined resonance. The interdigitated transducers (IDTs), consisting of 30 pairs of nanostrips, are located in the center of the NPM resonator for stimulating the phononic resonance. On the two sides of the IDTs, acoustic reflectors formed by 20 nanostrips are used to enhance the resonance. The pitch of the nanostrip array is set to 500 nm, which leads to an acoustic wavelength ($\lambda$) of 1 $\mu m$.

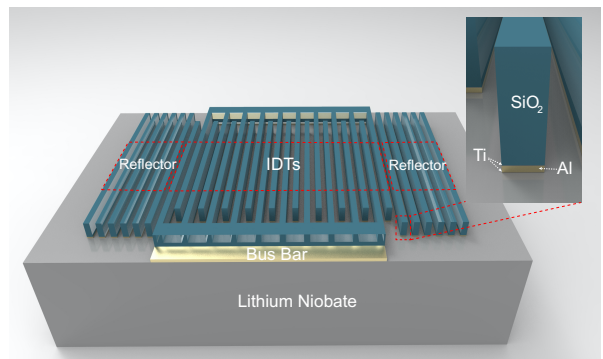

FIG. 1. NPM resonator with periodic high-aspect-ratio nanostrips covering the IDTs.

The device was fabricated by an electron-beam (e-beam) lithography based process. Titanium, aluminum and silicon dioxide were first deposited on the lithium niobate substrate. A complementary resist pattern of the device was then produced by e-beam lithography. After that, chromium etching mask was formed by e-beam evaporation and lift-off process. Reactive ion etching (RIE) was then used to remove the stacked materials between the high-aspect-ratio nanostrips. At last, conventional photolithography and another RIE were used to remove the silicon dioxide above the bus bars of the IDTs to expose the metal. Extra aluminum lines were deposited and patterned to extend the connection to the bus bars for external electrical connection. A scanning electron microscope (SEM) image of the fabricated device is shown in Fig. 2a.

Fig. 2b shows a cross-section of the high-aspect-ratio nanostrips obtained by focused ion beam (FIB). False color overlays are here used to identify the different materials. The green overlay shows the platinum deposited during the FIB cross-sectioning for protecting the top surface of the device. The metal electrodes (Ti and Al), silicon dioxide and lithium niobate are marked in yellow, blue and red, respectively. The nanostrip shows a rectangular profile with a slightly shrunk top and expanded bottom.

To obtain the resonance behavior of the NPM resonator, its scattering parameters were measured by a network analyzer. The impedance of the NPM resonator was then calculated by:

$$Z = Z_0 \frac{1+S_{11}}{1-S_{11}} \quad (1)$$

where $S_{11}$ is the return loss and $Z_0 = 50\,\Omega$ is the characteristic impedance. The corresponding impedance curves under 10, 20, and 30 dBm stimulation power are shown in Fig. 3a. The minimum and maximum of the impedance amplitude correspond to the resonance and anti-resonance points of the resonator. At 10-dBm stimulation power, the resonance frequency ($f_s$) and anti-resonance frequency ($f_a$) are equal to 1.422 GHz and 1.453 GHz respectively. The corresponding phase velocity ($v = f_s\lambda$) of the NPM wave is thus 1422 m/s, which is much lower than the velocity of the Rayleigh wave ($v_r = 3988\,m/s$). This reduced velocity is because of the slowing-down effect induced by the high-aspect-ratio nanostrips[13, 14] due to the hybridization between the surface acoustic wave and the elastic resonance of the high-aspect-ratio nanostrips. To reveal this hybridization, the mode shape of the NPM resonator in one wavelength (unit cell) was simulated using the finite element method (FEM) as shown in Fig. 3b. In contrast to conventional SAW resonators, in which the particle displacement is the largest on the substrate surface, the NPM resonator exhibits the largest particle displacement within the high-aspect-ratio nanostrips. The resonance frequency of the device is thus mostly determined by the geometrical dimensions and by material composition of the high-aspect-ratio nanostrips.

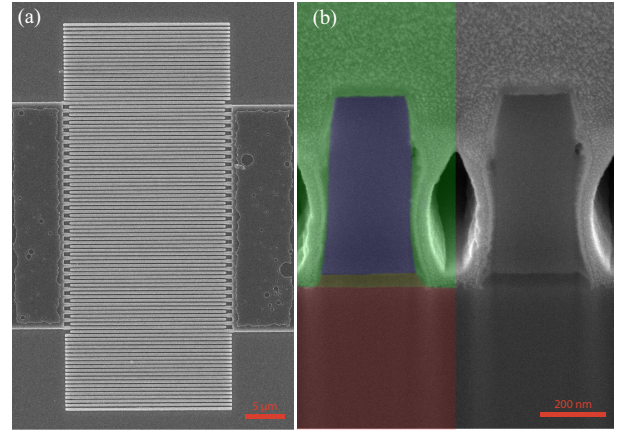

FIG. 2. (a) Top view SEM image of the NPM resonator (b) Cross-sectional view of the high-aspect-ratio electrodes.

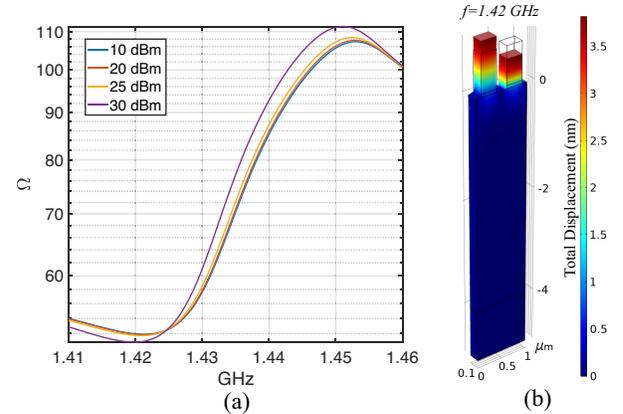

FIG. 3. (a) Impedance of a NPM resonator at different stimulation power. (b) Unit cell mode shape (enlarged displacement) of the NPM resonator at its resonance frequency.

To investigate the effect of stimulation power on the coupling induced nonlinearity, we measured the impedances of the NPM resonators under various stimulation power. Fig. 3a shows that the impedance curves shifts to the left with the increase of stimulation power, which represents a decrease of resonance frequency. This phenomenon is not observed in conventional thin-electrode SAW resonators with the same wavelength that were found to operate in linear regime under the same stimulation power. The measured frequency shift ($\Delta f = f - f_{10dBm}$) of the NPM resonators in the power range of 10 dBm to 30 dBm is shown in Fig. 4a. The four dashed lines represent the measurements of four devices fabricated in the same batch, which shows a good repeatability. As the frequency shift is negative, $|\Delta f|$ is shown in the figures for clarity. It can be seen that $|\Delta f|$ remains constant when the stimulation power is below 17 dBm. It then increases slowly until the power reaches 25 dBm. An accelerated frequency shift is then observed between 25 dBm to 30 dBm. The impedance at higher stimulation power is not measured as it exceeds the receiver power limit of the network analyzer. In a linear power scale shown in Fig. 4b, we found that the mean frequency shift of the four devices fits very well with a quadratic regression ($R^2 = 0.9996$) :

$$|\Delta f| = 0.5344P^2 + 1.274P - 0.0179 \quad (2)$$

where $\Delta f$ and $P$ are the frequency shift in MHz and stimulation power in Watt, respectively.

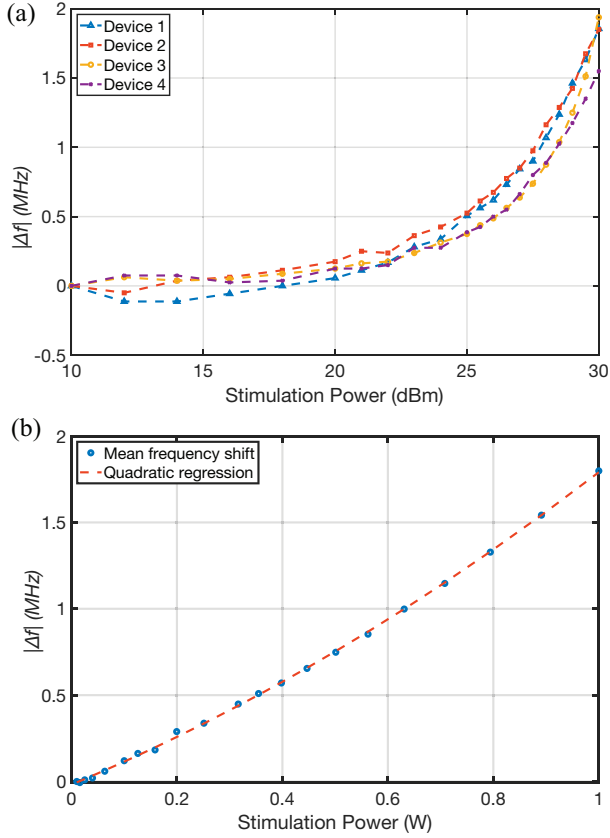

FIG. 4. (a). Power-dependent frequency shift of four NPM resonators. (b). Mean-frequency shift of the four devices and its quadratic regression.

To investigate if the power-dependent frequency shift also exists in larger scale electrodes, we fabricated an NPM resonator with a ten-time larger wavelength ($\lambda = 10\ \mu m$) for comparison. Experimental results showed constant resonance frequency in the stimulation power range of 10 dBm to 30 dBm, which means the 10-$\mu m$ $\lambda$ NPM resonator remains in the linear regime under the tested stimulation power range.

Nonlinearity of individual nanomechanical resonators could originate from various mechanisms including geometrical nonlinearity, material nonlinearity, actuation nonlinearity, etc[15]. The quadratic relationship between the stimulation power and frequency shift of the NPM resonator suggests its nonlinearity could possibly be attributed to geometrical nonlinearity[16]. However, a detailed analysis rules out this possibility. Geometrical nonlinearities of mechanical resonators appear when the displacement amplitude is sufficiently large compared to the device dimensions[15, 17]. In order to evaluate this effect in our device, an optical characterization of the mechanical vibration is achieved. Using a scanning laser heterodyne interferometer set-up inspired by Kimmo et. al.[18], we measured the out-of-plane displacements at two locations (Point A and Point B) on the nanostrips of an NPM resonator as shown in Fig. 5. It can be seen that the displacement amplitude increases linearly with the square root of the power until 400 mW. This is an expected result as the square of the displacement corresponds to the mechanical energy in the system, which is proportional to the electrical power of the stimulation. However, the displacement amplitude saturates, then decreases slightly after the stimulation power exceed 400 mW. As the geometrical nonlinearity relies on the increase of displacement amplitude to produce larger frequency shift, the saturation of amplitude should lead to a saturation in frequency shift, which is not observed in the experimental results (Fig. 4a). Another method to analyze the possibility of geometrical nonlinearity of the NPM is to consider the NPM as an array of singly-clamped beams. The geometrical nonlinearity of singly-clamped beam generally occurs in flexural modes when the displacement is comparable to the beam thickness[19]. The large displacement stretches the beam and induces extra tension within the beam, which in turn causes the frequency to change. However, the nanostrips constituting the NPM work in compression mode and have a displacement amplitude much smaller than their thickness. Moreover, geometrical nonlinearities usually results in a hardening effect of the fundamental mode that increases the mechanical resonator's resonance frequency[15, 16]. In contrast, the NPM exhibits a resonance frequency decrease with the increase of stimulation power that corresponds to a softening effect.

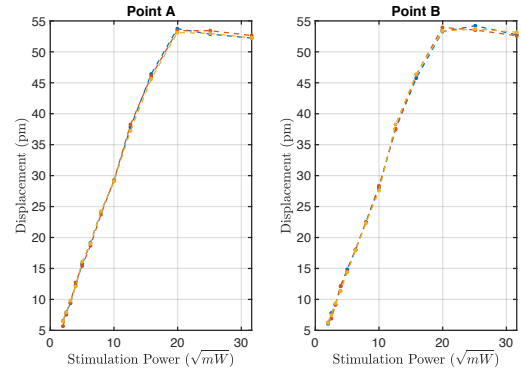

FIG. 5. Measurement of out-of-plane electrode displacement at two different locations on the IDTs of a NPM resonator.

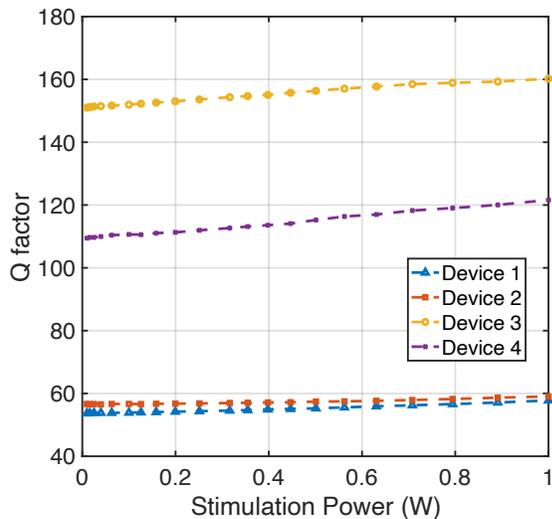

FIG. 6. Quality of the four NPM resonators at different stimulation powers.

It is previously reported that coupling induced by a covering graphene membrane on a phononic crystal can result in nonlinearity[11]. The NPM resonator actually share a similar mechanism in which the coupling between adjacent nanostrips is through the substrate instead of an external membrane. The resonance frequency decrease can be explained by the softening effect in the near surface of the substrate. This is because the increased coupling between adjacent nanostrips at high stimulation power improves the energy confinement along the substrate surface and thus softens this region. Because the improved energy confinement and enhanced coupling along the surface at high stimulation power, a larger ratio of the total elastic energy is stored in the near surface of the substrate, which means less energy is allocated to the nanostrips. This explains why a saturation of the nanostrip vibration amplitude is seen at high stimulation power (Fig. 5). The softening mechanism is also supported by the quality factor measurement as shown in Fig. 6: the stimulation–power-induced softening results in a decrease of the acoustic wave velocity in the surface layer of the substrate, which in turn reduces the acoustic radiation into the bulk substrate and improves the quality factor.

In summary, nonlinear effects causing a resonance frequency downshift in a nanostrip phononic metasurface operating at 1.42 GHz were observed. The measured frequency shift fits very well in a quadratic regression with $R^2$ reaching 0.9996. Comparative study using a 10-$\mu m$ wavelength NPM resonator with the same design shows a constant resonance frequency under the same stimulation power range, which means the device the has to be sufficiently small to exhibit the coupling-induced nonlinearity with stimulation power smaller than 1 Watt. Analysis shows that the nonlinearity is most likely originated from the coupling between adjacent nanostrips rather than by geometrical nonlinearity. By increasing the stimulation power, the coupling between the nanostrips is enhanced, which results in a softening effect of the substrate surface. The softening effect not only reduces the resonance frequency but also enhances the energy confinement along the substate surface and thus improves the quality factor. The power-dependent coupling in the NPM demonstrated a reliable method for the implementation of controllable nonlinearity in GHz phononic devices, which can enrich the mechanisms for the manipulation of surface acoustic waves at high frequencies.

**Acknowledgement**

This work is funded by NPRP grant No. NPRP10-0201-170315 from the Qatar National Research Fund (a member of Qatar Foundation). This work is also supported by the EIPHI Graduate School (contract "ANR-17-EURE-0002") and the French RENATECH network with its FEMTO-ST technological facility. The findings herein reflect the work, and are solely the responsibility of the authors.

**Data Availability**

The data that supports the findings of this study are available within the article.